\documentclass[12pt]{iopart}
\usepackage{iopams}
\usepackage{hyperref}
\usepackage{fullpage}
\usepackage{graphicx}
\usepackage{textcomp}

\usepackage{xcolor}
\definecolor{trueblue}{rgb}{0.0, 0.45, 0.81}
\definecolor{crimsonglory}{rgb}{0.75, 0.0, 0.2}

\usepackage{hyperref}
\hypersetup{colorlinks,linkcolor={trueblue},citecolor={crimsonglory},urlcolor={crimsonglory}}

\begin{document}

\title{Tunable dichroism and optical absorption of graphene by strain engineering}

\author{M. Oliva-Leyva and Gerardo G. Naumis}

\address{Depto. de F\'{i}sica-Qu\'{i}mica, Instituto de F\'{i}sica, Universidad Nacional Aut\'{o}noma de
M\'{e}xico (UNAM). Apdo. Postal 20-364, 01000, M\'{e}xico D.F.,
M\'{e}xico.}

\eads{\mailto{moliva@fisica.unam.mx},
\mailto{naumis@fisica.unam.mx}}
\date{\today}

\begin{abstract}

We theoretically study the transmittance for normal incidence of linearly polarized light between two media
separated by a strained graphene monolayer. We analytically characterize the degree of dichroism and the transparency of graphene
as a function of an arbitrary uniform strain and the incident polarization. We discuss how measurements of dichroism and transparency
for two different polarization directions can be used to determine the magnitude and direction of strain. Ours findings result in 
very useful tools to tune the graphene absorption by mechanical strain, as well as to design nano-devices to determine crack 
propagation in materials.

\end{abstract}



\maketitle

\section{Introduction}

Strain engineering has been widely used as an effective tool to improve the technological functionality of graphene. Its electrical,
chemical and optical properties are highly sensitive to mechanical deformation because of a unusual interval of elastic response. 
Graphene can withstand a reversible stretching up to $20\%$ \cite{Lee08}. 

Due to its relevance for graphene-based electronics, 
the band-gap opening in graphene is among the most investigated implications induced by deformations or electromagnetic 
fields \cite{Pereira09a,Ni08,Colombo10,Naumis08}.
Even though a uniaxial strain can be used to achieve the gap opening \cite{Pereira09a,Ni08}, theoretical studies have predicted that a combination of strains 
results in more accessible set-ups \cite{Colombo10}. Moreover, nonuniform deformations induce striking Landau levels due to effective pseudo-magnetic fields \cite{Guinea10a,Levy,Salvador13,Zenan14}.
Perfect graphene has a low piezoresistive sensitivity, however,
the graphene-based strain sensors are a promising field in nanotechnology \cite{Bae13,Shou13,Zhao13}. In nano- and micro-electromechanical systems, the  Casimir interaction
is a unwanted problem. In recent study, it was theoretically investigated how the strain modifies the force Casimir in graphene-based systems \cite{Phan14}.
On the other hand, a simple approach has been reported to control the chemical reactivity of graphene \cite{Bissett13}. The application of strain, 
via stretching of the supporting flexible substrate, produces impressive increases in the rate of reactivity \cite{Bissett13}.

Otherwise, the concept of strain engineering has been experimentally extended to the optical domain in recent works \cite{Pereira14,Bin14}. Unstrained graphene (undoped)
has a transparency of around  $97.7\%$ over a broad band of frequencies \cite{Nair08}. The origin of this remarkable feature, defined by fundamental
constants, is ultimately a consequence of graphene's unique electronic structure. Needless to say, another strain effect is the anisotropy in 
the electronic dynamics \cite{Our13}, which is traduced in an anisotropic optical conductivity \cite{Our14,Our14C}. Such effect for strained graphene yields a modulation of the 
transmittance as a function of the polarization direction. From a theoretical viewpoint, this modulation of the transmittance has been only quantified
in the case of a uniaxial strain \cite{Pereira10,Pellegrino09,Pellegrino10,Pellegrino11}. However, nowadays there are novel methods for applying biaxial strain in a controlled manner, even without 
the need for bending the substrate \cite{Shioya14}. So, a more general theoretical characterization of the strained-graphene transparency is needed.

In this paper, we quantify the modulation of the transmittance for graphene under an arbitrary uniform strain (e.g., uniaxial, biaxial, and so forth), 
as a function of the polarization direction. Also, we characterize the degree of polarization rotation as a function of strain and  
polarization direction. These results are useful to tune in effective manner the optical absorption of graphene, and hence, can be potentially
utilized towards novel optical detectors, sensors, and photovoltaics.

\section{Electromagnetic scattering problem}

\begin{figure}
\includegraphics[width=\textwidth]{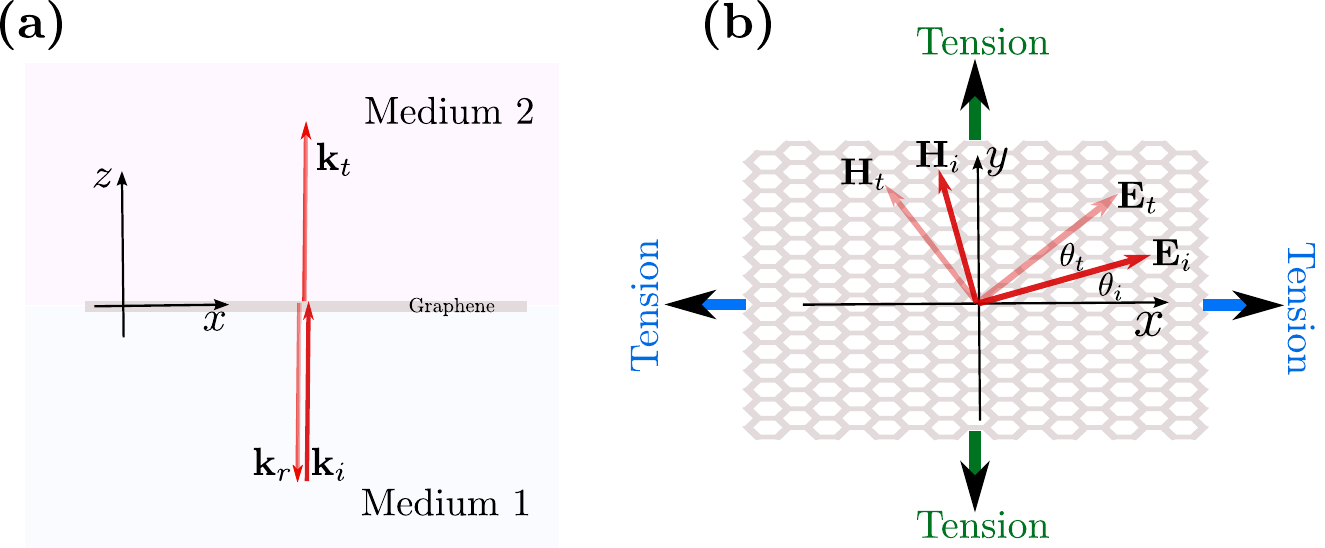}
\caption{(Colour online) (a) Front view of the scattering geometry for normal incidence between two media with strained graphene
separating them. (b) Schematic representation of strain-induced dichroism as seen from above the graphene sheet. 
The electromagnetic fields lie in graphene plane.}\label{fig1}
\end{figure}

We concentrate on the transmittance for normal incidence of linearly 
polarized light between two media
separated by a graphene sheet which is uniformly strained, as shown in Figure \ref{fig1}(a). 
We assume that the media are characterized by the 
electrical permittivities $\epsilon_{1,2}$ and the magnetic permeabilities $\mu_{1,2}$. From Figure \ref{fig1}(b),
we observe that the electric ($\bi{E}$) and magnetic ($\bi{H}$) fields 
belong to the graphene plane while the incident and transmitted polarizations $\theta_{i,t}$ are
measured with respect to the laboratory axes $xy$. 

For this scattering problem, one can set down immediately the boundary conditions
for the electromagnetic fields, which can be
written as \cite{Jackson},
\begin{eqnarray}
\bi{E}_{t}-\bi{E}_{i}-\bi{E}_{r}=0, \label{BC1}\\
\bi{n}\times (\bi{H}_{t}-\bi{H}_{i}-\bi{H}_{r})=\bi{J}, \label{{BC2}}
\end{eqnarray}
where $\bi{J}$ is the surface current density and $\bi{n}=(0,0,1)$. The electrical and magnetic field
on each media are related by 
\begin{equation}
 \bi{H}=\sqrt{\frac{\epsilon}{\mu}}\frac{\bi{k}\times\bi{E}}{k}, \label{HE}
\end{equation}
whereas Ohm's law reads
\begin{equation}
 \bi{J}=\bar{\bsigma}(\omega)\cdot\bi{E}_{t},\label{Ohm}
\end{equation}
with $\bar{\bsigma}(\omega)$ being the frequency-dependent conductivity tensor of graphene. Under uniform strain, 
$\bar{\bsigma}(\omega)$ is given by \cite{Our14,Our14C}
\begin{equation}\label{AC}
 \bar{\bsigma}(w)=\sigma_{0}(w)(\bar{\bi{I}}-2\tilde{\beta}\bar{\bepsilon}+\tilde{\beta}\Tr(\bar{\bepsilon})\bar{\bi{I}}),
\end{equation}
where $\bar{\bi{I}}$ is the $2\times2$ identity matrix, $\sigma_{0}(w)$ is the conductivity of unstrained graphene, $\bar{\bepsilon}$ is the strain tensor and $\tilde{\beta}\simeq2.37$ is 
related to Gr\"{u}neisen parameter. It is easy to see that an isotropic strain, $\bar{\bepsilon}=\epsilon\bar{\bi{I}}$, does not
affect the conductivity whereas an anisotropic strain yields an anisotropic conductivity. 

Combining (\ref{BC1})-(\ref{Ohm}), we obtain
\begin{equation}\label{Di}
 \bi{E}_{i}=\frac{1}{2}\sqrt{\frac{\mu_{1}}{\epsilon_{1}}}
 \left(\left(\sqrt{\frac{\epsilon_{1}}{\mu_{1}}}+\sqrt{\frac{\epsilon_{2}}{\mu_{2}}}\right)\bar{\bi{I}}+\bar{\bsigma}\right)\cdot\bi{E}_{t}.
\end{equation}

Equations~(\ref{AC}) and (\ref{Di}) shows how the strain-induced asymmetry of the conductivity tensor results in certain degree
of dichroism \cite{Pereira10}. Note that only for an isotropic conductivity (isotropic strain) the vectors $\bi{E}_{i}$ and
$\bi{E}_t$ are collinear and then the dichroism disappears. 

Now from (\ref{Di}) it is straightforward to write the transmittance as
\begin{equation}\label{T}
 T(\theta_{i})\approx T_{0}
 \left(1-\frac{2\sqrt{\mu_{1}\mu_{2}}}{\sqrt{\epsilon_{1}\mu_{2}}+\sqrt{\epsilon_{2}\mu_{1}}}\bi{e}_{i}\cdot\Re\bar{\bsigma}\cdot\bi{e}_{i}\right), 
\end{equation}
where $\bi{e}_{i}=(\cos\theta_{i},\sin\theta_{i})$ and $T_{0}$ is the transmittance for normal incidence between
two media in absence of the graphene interface. The term, $\bi{e}_{i}\cdot\Re\bar{\bsigma}\cdot\bi{e}_{i}=\Re[
\bar{\sigma}_{xx}\cos^{2}\theta_{i}+\bar{\sigma}_{yy}\sin^{2}\theta_{i}+\bar{\sigma}_{xy}\sin2\theta_{i}]$, shows
how an anisotropic absorbance yields the periodic modulation of the transmittance as a function of the polarization direction $\theta_{i}$.

In order to illustrate such effects, dichroism and modulation of $T(\theta_{i})$, let us assume both media to be vacuum ($\epsilon_{1,2}=\epsilon_{0}$ and $\mu_{1,2}=\mu_{0}$) and that the
graphene is at half filling, i.e., the chemical potential equals to zero. In this case, for infrared and visible frequencies, $\sigma_{0}(w)$ is frequency-independent
and is given by the universal value $e^{2}/(4\hbar)$. As a consequence, from (\ref{AC})-(\ref{T}) we obtain that the polarization 
angles $\theta_{i,t}$ are related by
\begin{equation}\label{DiR}
 \theta_{t}-\theta_{i}\approx\alpha\tilde{\beta}\bigl(\frac{\bar{\epsilon}_{yy}-\bar{\epsilon}_{xx}}{2}\sin2\theta_{i} + 
 \bar{\epsilon}_{xy}\cos2\theta_{i}\bigr)180^{\circ},
\end{equation}
whereas the transmittance results in,
\begin{equation}\label{TR}
 T(\theta_{i})\approx1-\pi\alpha\bigl(1-\tilde{\beta}(\bar{\epsilon}_{xx}-\bar{\epsilon}_{yy})\cos2\theta_{i} 
 - 2\tilde{\beta}\bar{\epsilon}_{xy}\sin2\theta_{i}\bigr),
\end{equation}
where $\alpha$ is the fine-structure constant. Now  it is easy to see the periodic modulations of the dichroism
and transmittance, with a period of $180^{\circ}$, which is a simple consequence of the physical equivalence
between the polarization angles $\theta_{i}$ and $\theta_{i}+180^{\circ}$\ for normal incidence of linearly 
polarized light. 

\begin{figure}[t]
\includegraphics[width=\textwidth]{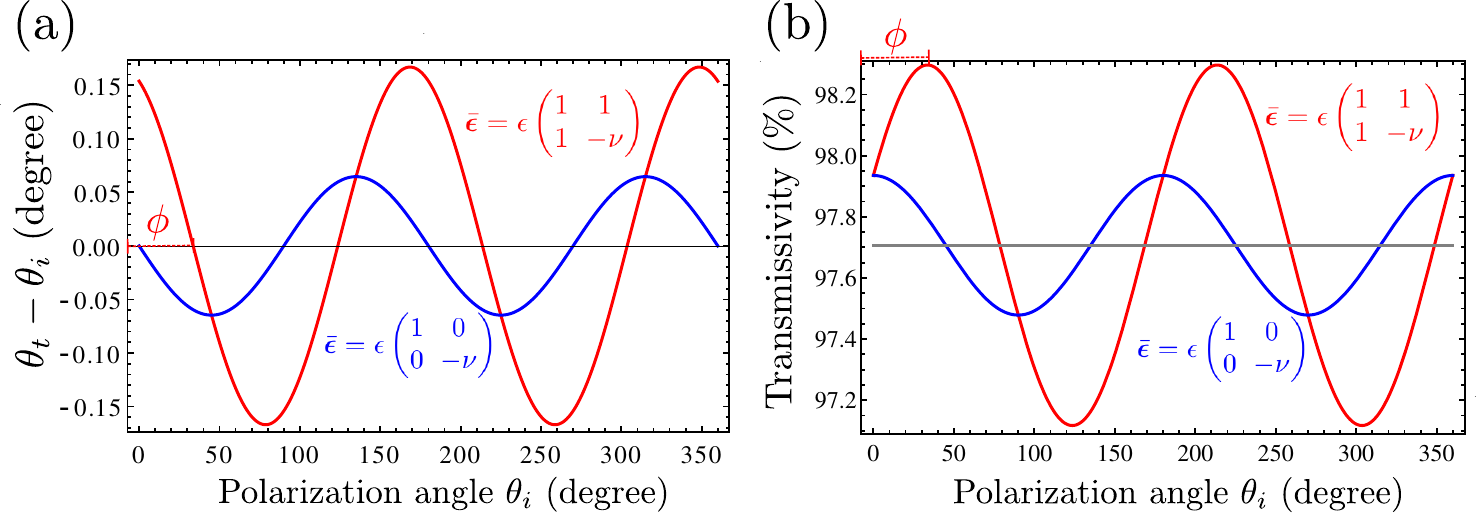}
\caption{(Colour online) (a) Rotation of the transmitted field and (b) transmittance as a function of the incident polarization angle for two
different deformations. The blue curves correspond to a pure uniaxial strain, while the red curves correspond to uniaxial-shear strain. 
The value $\epsilon=0.05$ was used for both deformations.}\label{fig2}
\end{figure}

\section{Discussion and Conclusions}

Let us explore first, the consequences on dichroism. In Figure~\ref{fig2}(a), we present the difference $\theta_{t}-\theta_{i}$ for two different deformations: a uniaxial deformation
and a uniaxial-shear deformation. The resulting modulations are out of phase with each other because the principal strain directions
of both deformations do not match.
For the uniaxial case considered, the principal strain directions match the laboratory axes $xy$, and $\theta_{t}-\theta_{i}$ displays a 
$\sin2\theta_{i}$-like modulation (see blue line). On the other hand, for the uniaxial-shear case, its principal strain directions do not match 
the laboratory axes $xy$, and thus $\theta_{t}-\theta_{i}$ results in a $\sim\sin(2\theta_{i}-\phi)$-like modulation (see red line), with 
$\tan\phi=2\bar{\epsilon}_{xy}/(\bar{\epsilon}_{xx}-\bar{\epsilon}_{yy})$. This behavior shows that 
the principal strain directions can be determined by measuring the polarization angles $\theta_{i}$ for which the incident and 
transmitted polarizations coincide. It is important to note that the polarization angles can be routinely measured with a precision of 
$0.001^{\circ}$, so that the  predicted modulation obtained from (\ref{DiR}) can be experimentally monitored. 

Finally, let us make some important remarks about our formula~(\ref{TR}) concerning transmittance.
First of all, one can see that (\ref{TR}) reduces to the universal value $1-\pi\alpha\approx97.7\%$, which is the transmittance of 
unstrained or isotropically strained graphene \cite{Nair08}. Likewise, (\ref{TR}) reproduces the previously reported modulation, 
$1-\pi\alpha(1-\tilde{\beta}(1+\nu)\epsilon\cos2\theta_{i})$,
for the case of a uniaxial strain along the \emph{x}-axis, where $\epsilon$ is the strain magnitude and $\nu$ is 
the Poisson ratio ($\nu\approx0.16$). In recent experiments, this periodic modulation of the transmittance for graphene
has been confirmed \cite{Pereira14}. In this experimental study, the strain magnitude $\epsilon$ was estimated from
the modulation amplitude measurements by means of $\epsilon\approx\triangle T/(2\pi\tilde{\beta}(1+\nu))$ and confirmed
from the Raman spectroscopy measurements \cite{Pereira14}.

Thus, (\ref{TR}) contains all previously studied limiting cases. Moreover, it can be used in the more
general scenario of biaxial strain. In fact, it provides a simple and reliable protocol to reconstruct the principal
axes of the strain tensor. The protocol goes as follows,

1) measure the transmittance at $\theta_{i}=0^{\circ}$,

2) measure the transmittance at $\theta_{i}=45^{\circ}$,

3) from  ~(\ref{TR}), 
\begin{equation}
\tan 2\phi=\frac{1-\pi\alpha-T(45^{\circ})}{1-\pi\alpha-T(0^{\circ})}
\end{equation}

4) then $\phi$ is just the angle between the laboratory axes $xy$ and the principal strain directions.


In conclusion, we calculated the dichroism and the transparency for normal incidence of linearly polarized light between two media
separated by a graphene monolayer under any arbitrary uniform strain. Our results contained some previously found particular cases.
Then we proposed a new protocol based in two simple transmittance measurements to reconstruct the applied strain field, and in 
particular, the principal axes of strain. Such protocol can be extremely useful to produce nano-sensors capable to detect the local
principal axis of strain, which are in fact determinant to determine crack propagation in different kind of materials. 
Also, it can serve to measure pseudo-magnetic fields associated with graphene.

\ack {We thank the DGAPA-UNAM project IN-102513. M. Oliva-Leyva
acknowledges a scholarship from CONACyT (Mexico).}

\section*{References}

\bibliographystyle{unsrt}
\bibliography{biblio_Transmittance}

\end{document}